\documentclass{article}

\usepackage[preprint,nonatbib]{neurips_2022}

\usepackage[utf8]{inputenc} 
\usepackage[T1]{fontenc}    
\usepackage{hyperref}       
\usepackage{url}            
\usepackage{booktabs}       
\usepackage{amsfonts}       
\usepackage{nicefrac}       
\usepackage{microtype}      
\usepackage{xcolor}         
\usepackage[pdftex]{graphicx} 
\usepackage{algorithm}
\usepackage{algorithmic}
 
\usepackage{amsmath}
\newcommand{\system}{\textsc{Hydra}}

\title{Hydra: A System for Large Multi-Model Deep Learning}

\author{
 Kabir Nagrecha\\
  Computer Science and Engineering\\
  University of California San Diego\\
  La Jolla, CA 92023 \\
  \texttt{knagrech@ucsd.edu} \\
  \And
  Arun Kumar \\
  Computer Science and Engineering\\
  University of California San Diego\\
  La Jolla, CA 92023 \\
  \texttt{arunkk@eng.ucsd.edu} 
}

\begin{document}

\maketitle

\begin{abstract}
Scaling up model depth and size is now a common approach to raise accuracy in many deep learning (DL) applications, as evidenced by the widespread success of multi-billion or even trillion parameter models in natural language processing (NLP) research.
Despite their success in DL research and at major technology companies, broader practical adoption of such large models among domain scientists and businesses is still bottlenecked by GPU memory limits, high costs of training or fine-tuning, and low GPU availability, even on public clouds. These resource challenges are further compounded by model selection needs: DL users often need to compare dozens of models with different hyper-parameter combinations and/or neural architectural design choices to suit their specific task and dataset. In this paper, we present \system, a system designed to tackle such challenges by enabling out-of-the-box scaling for multi-large-model DL workloads on even commodity GPUs in a highly resource-efficient manner.
\system~is the first approach to holistically optimize the execution of multi-model workloads for large DL models. We do this by adapting prior ``model-parallel'' execution schemes to work with scalable parameter offloading across the memory hierarchy and further hybridizing this approach with task-parallel job scheduling techniques.
\system~decouples scalability of model parameters from parallelism of execution, thus enabling DL users to train even a 6-billion parameter model on a single commodity GPU. It also fully exploits the higher speedup potential offered by task parallelism in a multi-GPU setup, yielding near-linear strong scaling and in turn, making rigorous model selection perhaps more practical for such models.
We evaluate end-to-end performance by fine-tuning GPT-2 for language modeling. We find that \system~offers between 50\% and 100\% higher training throughput than even the best settings of state-of-the-art industrial frameworks such as DeepSpeed and GPipe for multi-large-model training.
\end{abstract}

\section{Introduction}
\label{sec:introduction}
The high profile success of DL at big technology companies has led to high interest in adopting state-of-the-art DL models at smaller companies in the Web, enterprise, and healthcare sectors, as well as among domain scientists and in digital humanities. Large neural architectures such as Transformers and other so-called ``foundation models''~\cite{foundationmodels} now dominate NLP and have multiple billions of parameters, e.g., BERT-Large~\cite{devlinBert}, GPT-3~\cite{brown2020language}, and Megatron-LM~\cite{shoeybi2019megatron}. Interest in such large models is also growing in computer vision (e.g.,~\cite{vit}) and for tasks bridging NLP and tabular data~\cite{tabert}. Moreover, the popularity of transfer learning using base models provided by public libraries such as HuggingFace~\cite{huggingface} is powering a massive shift toward ``low-data large-model'' training setups~\cite{foundationmodels}. 
Alas, three key systems- and economics-related bottlenecks are impeding the adoption of such powerful models by DL users outside of big technology companies: (1) GPU memory capacity trailing DL model sizes~\cite{sohonilowmem}, (2) high computational/cost/energy footprints of GPU clusters, and (3) high demand for GPUs relative to supply, including on public clouds. Thus, ensuring \textit{overall resource efficiency}, as well as \textit{enabling DL users to make do with fewer GPUs and/or cheaper GPUs} is a pressing research concern to ensure that the potential of large-scale DL models are accessible to the many, not just the few.

The main approach practiced today to mitigate bottleneck (1) above (viz., GPU memory limits) is to \textit{partition} the model's neural computational graph across multiple GPUs to lower its memory footprint on each GPU. This form of execution, known as \textit{model-parallelism}, is increasingly popular~\cite{modelparallelism}. However, model-parallelism suffers from two fundamental issues.

First, sequential dependencies within the neural architecture causes resource idling (busy waiting) and thus, GPU underutilization. Figure 1(C) illustrates how devices can be blocked while waiting for intermediate data to be passed forward/backward by earlier stages of the model. However, this issue is mitigated to an extent by \textit{pipeline-parallelism}, which shuttles different batches of data through different stages of the model in parallel. 
Another technique known as \textit{tensor-parallelism}, which divides a model width-wise, can also help~\cite{flexflow}.
We explain more about these techniques in Section~\ref{sec:related}. Nevertheless, some significant amount of resource idling is still inevitable in such techniques if one must preserve correctness (i.e., no heuristic approximations).

Second, most DL users do not train just one model in a vacuum but rather do it as part of a larger multi-model execution scenario. Model selection needs such as tuning hyper-parameters and/or fine-tuning some layers of the network is needed to control the balance of overfitting and underfitting on a new task and dataset~\cite{understandingml}. That leads to multi-model execution. Multi-tenant clusters also see multiple models being trained together. In such scenarios, \textit{task-parallelism}, viz., a job scheduler assigning different models to different workers, helps raise throughput of execution. But pure model-parallelism works directly against task parallelism in such cases. Raising the per-model footprint to multiple GPUs reduces the number of tasks one can run in parallel on a given cluster and/or forces users to raise their cluster sizes by trying to get even more (expensive) GPUs.

\paragraph*{\textbf{Example.}} 
Consider a political scientist building a text classifier for sentiment analysis of tweets to understand polarization between gun rights and gun control supporters in the US. 
They download a state-of-the-art GPT-2 model from HuggingFace to fine-tune it. They decide to compare a few different learning rates and optimizers with a grid-search, leading to 48 different model configurations to train. Using an AWS P3 node in the N.~Virginia region that offers Tesla V100 GPUs, they first try to train one model on a single 16GB GPU (\$3.06/hr). Alas, the model's size causes out-of-memory (OOM) errors, with both PyTorch and TensorFlow crashing. So, they switch to 4-GPU node to train it in a model-parallel manner, costing \$12.24/hr. But then they realize that fine-tuning even for a few epochs could take multiple hours and grid search in a serial fashion (one model after another) would be too slow and take weeks. They consider manually overlaying task-parallelism on top of model-parallelism, costing them up to \textit{\$590/hr}. But AWS rate-limiting policies prohibits them from obtaining 192 GPUs, forcing them to either move up to a much more expensive GPU and/or suffer much longer runtimes. Anecdotally, these sorts of frustrations are now common among DL users. 

\textit{Overall, we observe that today's DL systems have a dichotomy of model-parallelism and task-parallelism for multi-large-model DL workloads. This leads to substantial resource idling and GPU underutilization, which in turn leads to higher runtimes, costs, and energy footprints.}

In this paper, we start with a simple insight: \textit{the above dichotomy is a false dichotomy and we devise an approach that enables simultaneous task-parallel and model-parallel training of large DL models.}
We note that a key issue with today's model-parallelism is that it forces users to get multiple GPUs simply to store a model in the aggregate multi-GPU memory. We remove that bottleneck from first principles by using a ``spilled'' execution scheme that enables model-parallel scalability without the need for multiple GPUs. We do so by automatically rewriting a full model into shards (or sub-models) and promoting and demoting such shards between GPU memory and DRAM. This allows us to support very large feedforward models (e.g. Transformers, CNNs, MLPs) on even just a single GPU, decoupling scalability from parallelism. We leave non-feedforward-architectures such as graph neural networks and recurrent nets to future work.

Building on top of our above style of model spilling, we devise a \textit{novel hybrid of task-parallelism and model-parallelism} we call Shard Alternator Parallelism (SHARP). SHARP offers the advantage of high throughput and compute-scalability (exploiting the higher degree of parallelism in multi-model workloads) of task-parallelism but does not suffer its disadvantage of needing to fit a full model into a GPU's memory. Likewise, SHARP offers the advantage of model-scalability of model-parallelism (not needing to fit a full model in a GPU's memory) but does not suffer its disadvantage of sequential dependencies leading to low throughput and poor compute-scalability.

We implement our above techniques into a system we name \system~on top of PyTorch. We offer it as an open-source library available under Apache License v2.0. We demonstrate \system's benefits for multi-large-model DL workloads by performing a grid search to fine-tune GPT-2 for language modeling on the WikiText-2~\cite{wikitext2} dataset (available under Creative Commons License) on two different GPU setups. We find that \system~enables us to surpass the state-of-the-art perplexity results, while offering 1.5-4.8X faster runtimes and between 50\% and 100\% higher training throughput compared to DeepSpeed~\cite{zeroDeep,zeroOpt,zeroInfinity} and GPipe~\cite{gpipe}, two state-of-the-art industrial-strength tools for large-model DL training. We also show that \system~is able to scale up DL model sizes on a commodity GPU by training a 6-billion parameter model on a 16GB GPU, while the other tools crash at much smaller model sizes.

\section{System Details}
\label{sec:details}
We now describe the interface and implementation details of \system. \system~is provided to users as an open-source library available under the Apache License. Using \system~is relatively simple --- it acts as a drop-in replacement for the typical PyTorch training loop, acting directly on top of PyTorch modules. This eases integration and adoption considerably.

\begin{figure}
\label{fig:sharp}
  \centering
  \includegraphics[width=\linewidth]{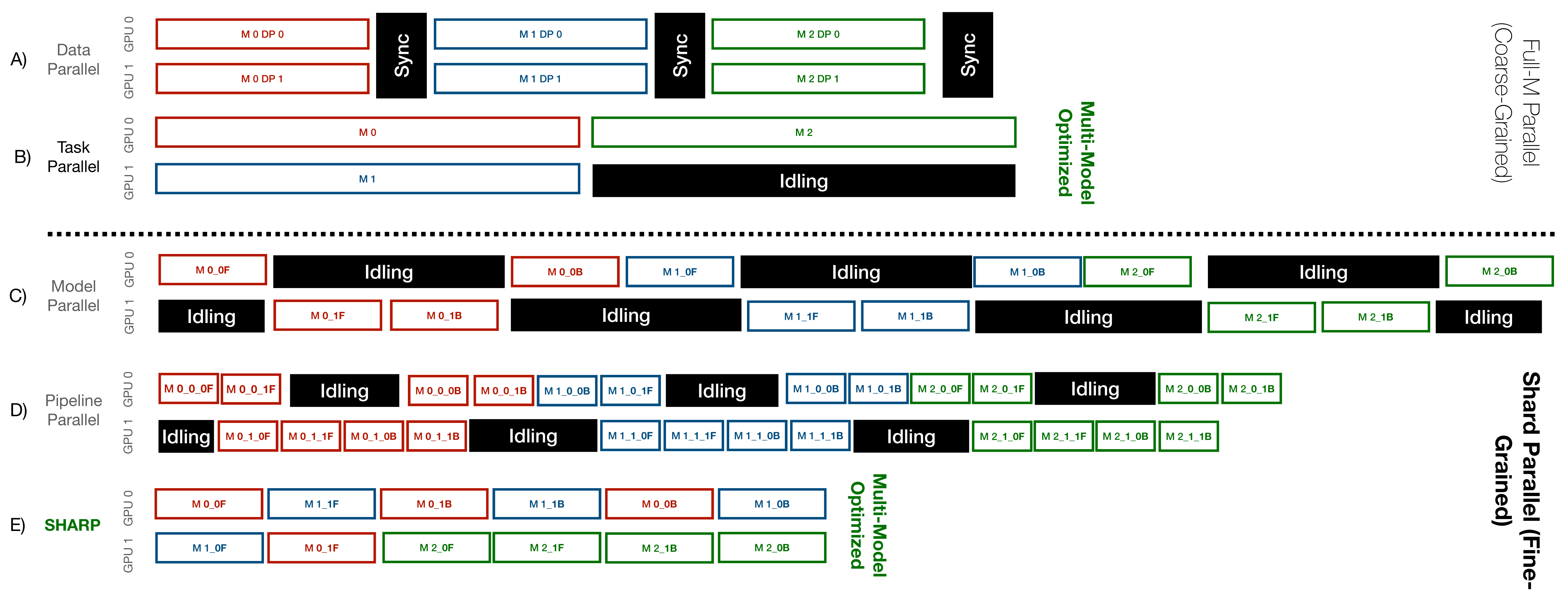}
  \caption{Simplified illustration of training three models for a single minibatch with various techniques. We use $\{model\}\_\{shard\}$ format to describe fine-grained execution of shards; the additional postfix $\{microbatch\}$ is used for pipeline parallelism. The suffix F or B indicates forward or backward pass. With SHARP, we exploit the efficiency of task parallelism and combine it with the scalability and fine-grained optimization of model parallelism to minimize runtimes and idling.}
\end{figure}

\subsection{Interface}

\system~takes as input a list of PyTorch models, PyTorch dataloaders, and training job specifications (e.g. loss functions and hyperparameters), then orchestrates execution. \system~automatically generates partitioning strategies and execution schedules with minimal user input. We provide more detail on the API usage in supplementary materials. Pretrained model libraries such as HuggingFace integrate easily with minimal development overhead.

\subsection{Partitioner}
\label{sec:partitioner}
\system~begins by analyzing the memory footprint of each user-provided model with respect to GPU memory bounds. We introduce a simple automated partitioner, described in Algorithm~\ref{alg:partitioning}, that runs a sample minibatch through the model in a pilot pass and introduces ``cut-points'' when GPU memory is overloaded. In this way, \system~can shard a model into subgraphs of the original architecture's neural computational graph. These partitions are \system's equivalent of model-parallel shards. The user is then provided with logging output informing them where their model was partitioned. So, in future runs with the same architecture they can reuse the same partitioning directly without a new pilot pass. The pilot pass also provides us with runtime statistics for future use.

\begin{algorithm}[tb]
 \caption{Dynamic model partitioning algorithm.}
  \label{alg:partitioning}
 \begin{algorithmic}
 \STATE {\bfseries Input:} Model as a sequence of $m$ layers $L$; data mini-batch $B$; GPU $G$
  \STATE {\bfseries Output:} Array of partition indices $A$
 \STATE Append 0 to $S$
 \FOR{$i=0$ {\bfseries to} $m-1$}
 \STATE Place $L[i]$ and $B$ on $G$
 \STATE $B'$ $\leftarrow$ Forward pass through $L[i]$ with $B$
 \STATE$T$ $\leftarrow$ New tensor with same shape as $B'$
 \STATE  Backpropagate $T$ through $L[i]$ without releasing memory
  \IF{$G$ out of memory}
   	\STATE Append $i$ to $S$
   	\FOR{$j=0$ {\bfseries to} $i-1$}
		\STATE Release all memory consumed by L[j]
		\STATE Append $i$ to $A$
	\ENDFOR
   \ENDIF
   \ENDFOR
  \end{algorithmic}
\end{algorithm}

Note that our algorithm assumes that the model graph is a chain architecture (sequence of layers). This structure suffices for most large-model architectures such as Transformers, CNNs, and MLPs. Recurrent and graph neural networks are out of scope for \system. In model parallel execution, the generated shards would be placed on different GPUs to arrange the model across a network of devices. However, as we previously discussed, this execution strategy drives up compute requirements and minimizes the degree of task parallelism we can employ in multi-model workloads. As such, we now look for a novel execution strategy that will enable us to run our shards even if there is only one GPU.

\subsection{Spilling}
\textit{Spilling} is a memory hierarchy utilization technique from the relational database management space that enables large data to be processed with low memory. We adapt this technique to enable scalable and flexible large-model execution. Essentially, we ``chunk'' model execution into sharded stages according to a partitioning scheme, then sequentially promote and demote model partitions and intermediate data between DRAM and GPU memory. Figure 2(A) illustrates. Because spilling directly replicates model parallel execution, and model parallel execution is known to be mathematically equivalent to standard execution~\cite{modelparallelism}, spilling is also mathematically equivalent to standard execution.

Our approach bears some resemblance to previous offloading designs explored in works such as ZeRO-Infinity~\cite{zeroInfinity} and SwapAdvisor~\cite{huang2020swapadvisor} but generalizes the concept further to enable flexible multi-model scheduling. We discuss the differences in depth in Section~\ref{sec:related}. 

Each shard is loaded to GPU memory twice, once during the forward pass and once during the backward pass. Backpropagation requires reuse of activations generated during the forward pass, but this would substantially increase CPU-GPU communication overheads. Instead, we make use of gradient checkpointing~\cite{chen2016training}, saving activations at shard boundaries and recomputing in-shard intermediates during the backward pass. A similar approach was used to reduce memory bloat in GPipe~\cite{gpipe}. 
Even with checkpointing, communication latencies can be substantial. In our initial evaluations, naive spilling incurred a 3X overhead versus model parallel execution using fast GPU-GPU interconnects. To mitigate this, we use \textit{double buffering}, a latency-hiding technique, to overlap communication with compute by prefetching shard parameters to a buffer space on the GPU while a different shard is still executing. This buffer space can be relatively small, as model parameters tend to be less than 10\% of the model's overall memory footprint during execution~\cite{gpipe, zeroOpt, l2l}. Empirical evaluations training GPT-2 on a single Tesla V100 with our approach took only 15\% longer than model parallelism on 4 of these GPUs. Likewise, spilling on a cheaper K80 GPU was only 80\% slower than model parallel execution on 8 of these GPUs. The slowdown factor is dependent on many things, including CPU-GPU interconnect speed, GPU-GPU interconnect speed, GPU memory, and GPU's processing speed.

Spilling enables us to scale to larger-than-GPU-memory models even on a single GPU as Figure 2(B) illustrates. More critically for \system, being able to train large models even with just one GPU via sharding and spilling enables us to exploit task parallelism to its fullest extent. Since task parallelism offers linear strong scaling, \system~can hide the slowdowns we noted earlier in multi-GPU environments, surpassing model parallelism. Thus, spilling is beneficial for both the low-resource user (by enabling scaling) and for the high-resource user (by enabling maximal parallelism).

\begin{figure}
 \label{fig:spilling}
  \centering
  \includegraphics[width=0.9\linewidth]{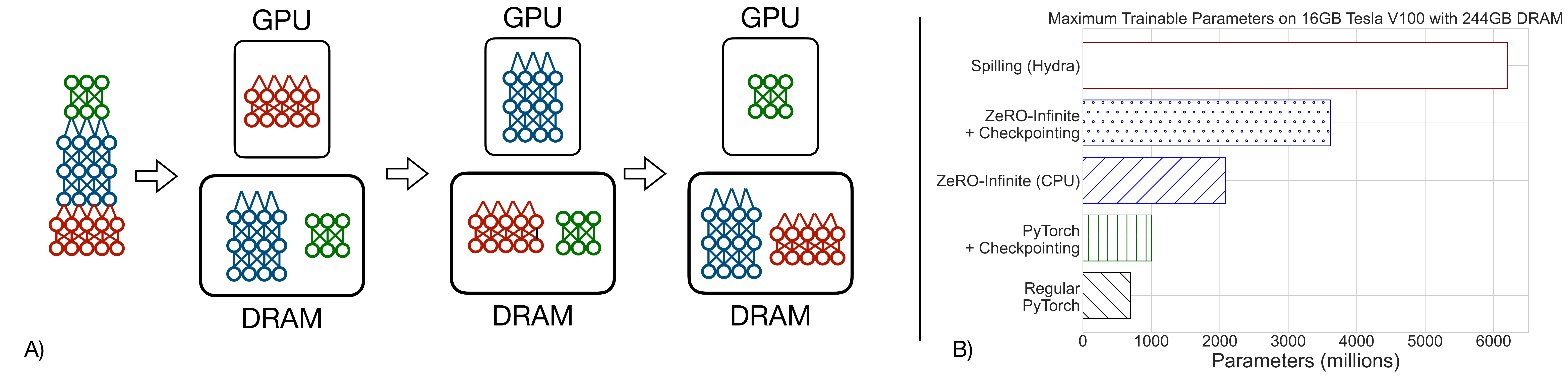}
  \caption{A) Temporal schematic of a spilled forward pass. B) Demonstration of spilling's scalability versus popular techniques for single-GPU large-model training. We train scaled up GPT-2 models using a batch size of 1 and context length of 512 to explore the maximum trainable model size using different DL systems.}

\end{figure}

\subsection{Shard Alternator Parallelism}

Shard Alternator Parallelism, or SHARP for short, is our hybridization of spilling with task parallelism. We identify several desirable characteristics of task parallelism: zero-synchronization parallel execution and linear strong scaling when there are more tasks than processors. These are qualities we wish to preserve in our hybridization. However, we also identify a drawback: poor heterogeneous scheduling efficiency. When tasks have different execution lengths, task parallelism can only maximize parallelism for a limited period of time. Using fine-grained parallelism enables us to work around this issue so long as we have sufficient tasks to choose from. Figure 1 illustrates these tradeoffs.

With SHARP, we aim to combine the scalability and fine-grained optimization of model parallelism with the lower resource idling of task parallelism. Note that this is only possible through the flexibility of spilling. To make the most of fine-grained parallelism, we must understand how to \textit{schedule} multiple models at a fine-grained, sub-model level. This is especially critical for extreme scale models, where each individual minibatch potentially introduces hundreds of sharded execution tasks. After every shard completion, we must select a model to provide a shard for the newly freed device. Depending on the workload, the user could have one of several different scheduling objectives. In \textit{batched} multi-model jobs, such as model selection, which is our focus, individual model training latency is less critical than overall completion time of the workload. We formalize the scheduling problem with completion time as the objective as an MILP in supplementary materials.

Using an optimal solver such as Gurobi~\cite{gurobi} for this task is not practical given the sheer number of shard execution tasks in our setting (even in the millions). Instead, we look for a fast and easy-to-implement dynamic scheduler. Intuitively, we can identify two settings that our scheduler encounters when training batched multi-model jobs. Initially, the workload will likely be have more model-tasks than GPUs. It is easy to maximize throughput and utilization in this setting, as every processor can be kept busy with a model. Over time though, tasks will complete, and there will be fewer tasks than devices. Figure 1(B) illustrates how this reduces the upper bound of our achievable degree of task parallelism.

We can minimize time spent in this reduced-efficiency setting by completing all tasks at approximately same time. This is a known property of "longest-remaining-time", or LRTF, schedulers~\cite{osbook}. Unlike standard LRTF implementations which run tasks continuously with occasional pre-emptions, we treat each individual shard as its own atomic task with the time-cost being defined by the total \textit{model}'s running time. This maintains our desired scheduling behavior (even task completion times, maximal processor utilization) while fitting into the sharded nature of spilled execution. We name our dynamic greedy scheduler Sharded-LRTF. Empirical evaluations of Sharded-LRTF in our supplementary material demonstrate that Sharded-LRTF produces near-optimal execution times thanks to its fine-grained scheduling while incurring minimal scheduling overheads. With these techniques --- spilling, SHARP, and Sharded-LRTF --- \system~is able to optimize multi-large-model workloads holistically.

\section{Language Modeling Experiments}
\label{sec:experiments}
\textbf{Dataset and Workload.} We now evaluate \system's performance on real-world workloads. Language modeling is a core task in the NLP space and fine-tuning pretrained models for language modeling is a common workload. We use \system~to run a model selection job for fine-tuning open-source HuggingFace GPT-2 models~\cite{huggingface} on the WikiText-2 dataset~\cite{wikitext2}. The workload compares 12 different hyperparameter configurations, obtained by a grid search with 2 batch sizes \{16, 8\} and 6 learning rates \{0.0003, 0.0001, 0.00005, 0.00006, 0.00001, 0.00002\} inspired by real-world GPT-2 fine-tuning jobs~\cite{hellogpt,gpt2portugal,huggingfaceexample,rutransformers}. We use a context length of 512 tokens. We do not freeze parameters or take any steps to reduce the computational workload --- we want \system~to undergo the full load of training.

\begin{figure}
\label{fig:charts}
  \centering
  \includegraphics[width=\linewidth]{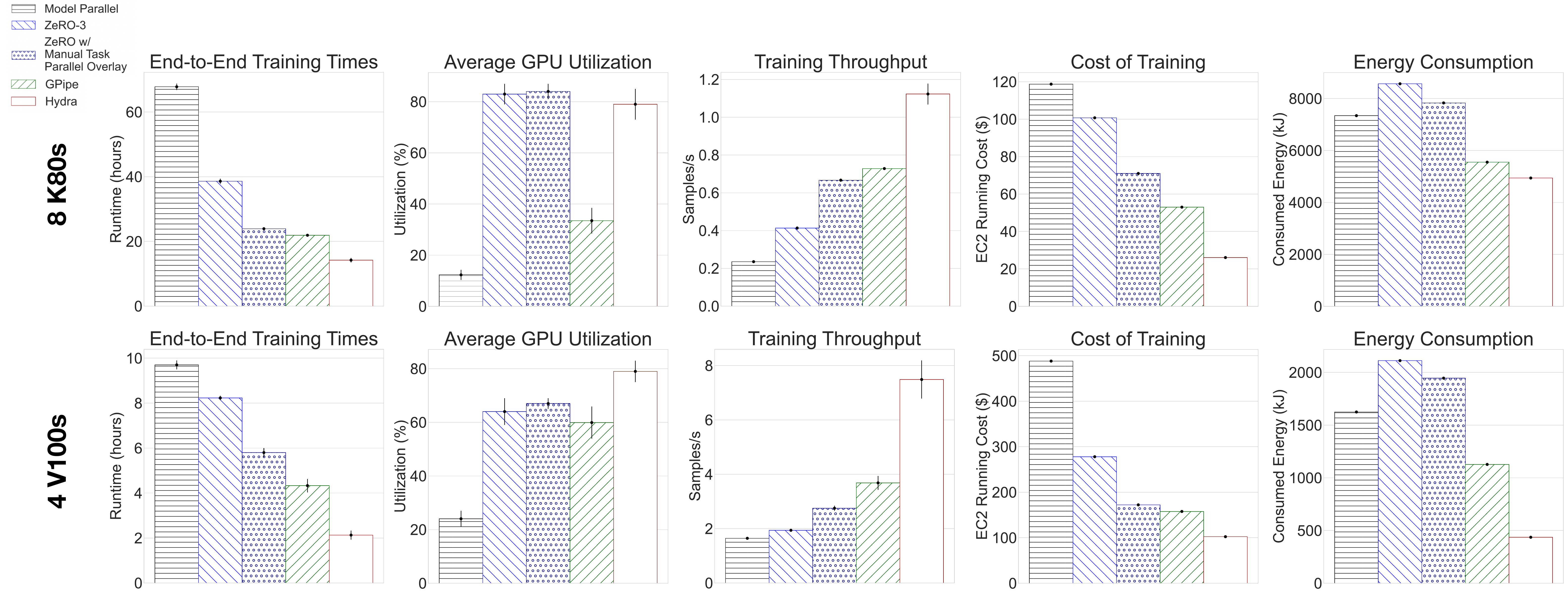}
  \vspace{-4mm}
  \caption{Critical statistics recorded from language-modeling GPT-2 model selection jobs. We run a manual hybrid task parallelism over ZeRO-3 to simulate manual user-set task parallelism. GPipe and naive model parallelism crash due to GPU memory errors when run on 4 K80s or 2 V100s for this job, so we are not able to overlay manual task parallelism for those two approaches.}
\end{figure}

We compare against two state-of-the-art industrial strength tools, ZeRO-3~\cite{zeroOpt,zeroDeep} provided through the Microsoft DeepSpeed library and a PyTorch implementation of GPipe~\cite{gpipe,torchgpipe}. Figure~\ref{fig:charts} shows our end-to-end performance benchmarked against those two tools, a manual task parallel hybrid on ZeRO-3, and base model parallelism. The sheer scale of large language models also raises concerns of energy consumption and running costs. So, we report on these metrics in addition to the traditional runtime/utilization numbers. One of our core aims in this paper is to demonstrate \system's usability on accessible hardware. Prior works demonstrating techniques for large-model training (e.g. Megatron~\cite{shoeybi2019megatron}, ZeRO-3) have generally focused on large-cluster, high-performance hardware configurations with expensive Nvidia DGX-2 servers. Our experiments are all run on AWS to create a reproducible environment. While \system's techniques could certainly be applied to larger, more powerful hardware configurations, we do not focus on these settings in this paper. 

\textbf{End-to-End Results.} We run two sets of jobs, one on 8 K80 GPUs, and another on 4 Tesla V100s.  In both settings, we find that \system~reports the lowest execution times, costs, and energy consumption. Compared to ZeRO-3, \system~is 2.5X faster out-of-the-box on 8 K80s and 3.5X faster on 4 V100s. Applying a manual task parallel overlay on top of ZeRO-3 improves its performance but it still falls behind \system's efficiency, demonstrating that our approach of integrating task parallelism from the ground-up outperforms top-down hybridizations. The closest competitor is GPipe, which makes use of a fast NVLink connector for GPU-GPU communication. \system's use of GPU-CPU-GPU communication with spilling should disadvantage it; yet \system~reports 50-90\% lower runtimes and comparable or up to 2X lower energy consumption. Naive model parallelism produces by far the worst performance, about 4X slower than \system~in both settings.

\begin{table}
\label{tb:accuracy}
\centering
\begin{tabular}{ c c c }
Model & Test Perplexity & Validation Perplexity\\ [0.5ex] 
 \hline\hline
  \textbf{GPT-2 (fine-tuned with \system)} & 15.17 & 15.69 \\    
 GPT-2 (Zero Shot) & 18.34 & 19.6 \\ 
 BERT-Large-CAS & 34.1 & 37.7 \\    
\end{tabular}
\vspace{1.5mm}
 \caption{Fine-tuned model accuracy compared to zero-shot GPT-2~\cite{radford2019language} and BERT-Large-CAS~\cite{wang2019cas}. We only fine-tune for one epoch, but \system~could be easily be use to run more extensive model selection jobs or even build new architectures.}
\end{table}

We initially planned to benchmark against Megatron-style 3D parallelism~\cite{narayanan2021efficient}, but based on their GitHub repository, we found that the only readily accessible implementation of 3D parallelism, provided by Microsoft's DeepSpeed library, is not yet usable for out-of-the-box model training~\cite{deepspeedissue1,deepspeedissue2} and is restricted to a limited set of training examples. Since our aim is to compare to typical large-model model selection options available to general DL users, Megatron-style 3D parallelism is not yet a practical candidate for comparison.

\textbf{Accuracy.} Table 1 compares the accuracy of our final model, with selected learning rate 0.0003 and batch size 8, to a few published examples. The full results of each configuration are available in supplementary materials. Please note that the aim of this experiment is \textit{not} to claim that we have a better model than GPT-2. This is not a fair comparison --- we are reporting against a \textit{zero-shot} version of GPT-2. Fine-tuning will naturally improve results. We only report accuracy to demonstrate that \system~can be used to produce state-of-the-art results and advance DL research and practice. 

\section{Related Work}
\label{sec:related}
Over the past several years, many systems have been released to support distributed execution for large-model training. Unlike prior approaches, \system~exploits task parallelism --- it is the \textit{first system to holistically optimize multi-large-model training}.

\textbf{Alternative approaches to model parallelism} (e.g. tensor parallelism) shard models in a more complex fashion, partitioning individual layers into pieces rather than dividing a model into shards. This increases complexity substantially but opens up more possibilities for parallel execution. Indeed, ZeRO~\cite{zeroDeep,zeroOpt} uses tensor parallelism in combination with data parallelism to offer higher training efficiency. We note, however, that tensor parallelism's complexity increases communication overheads and per-architecture implementation effort, especially when compared with the simplicity of spilling. In either case, these techniques are orthogonal to our goal of exploiting task parallelism in multi-model workloads. We leave it to future work to hybridize these techniques, but anticipate that communication challenges will be a challenge when combining tensor parallelism with spilling.

\textbf{Hybridizations between model parallelism and data parallelism}~\cite{zeroDeep,zeroOpt,zeroInfinity} are now widely used to improve large-model training performance. ZeRO, for example, combines the multi-GPU requirements of intra-layer (tensor) parallelism with the multi-GPU requirements of data parallelism, eliminating the memory bloat of traditional data parallelism. Empirical evaluations with ZeRO demonstrate that it offers substantially better performance than naive model parallelism along with better scaling. However, the communication overheads of data parallelism weigh heavily on its performance, especially when compared with zero-synchronization task parallelism. Moreover, data parallelism requires the user to treat a training hyperparameter (batch size) as a control for efficiency in addition to model accuracy, which can be problematic in model selection workloads. We note that ZeRO's data parallelism could be hybridized with \system~to address the most substantial weakness of task parallelism --- poor efficiency when there are fewer models than processors. We leave this additional hybridization to future, as explained further in Section~\ref{sec:improvements}. Model-task hybrids were initially proposed in a short abstract presented at a non-full length (2 page) venue~\cite{hydra_cerebro}. 
This paper expands those concepts into a complete problem setting with a full solution by fleshing out hybrid model-task parallelism along with a thorough empirical evaluation on real DL workloads.

\textbf{Pipeline parallelism} is one of the most popular modifications of model parallel execution. GPipe~\cite{gpipe} proposed using the sharded model as a staged-out pipeline, and shuttle mini-batch partitions, known as \textit{micro-batches}, through the model. While this increases utilization and throughput over model parallelism, the bi-directional nature of model execution (i.e. prediction and backpropagation), forces pipeline flushes between directional changes. Other works such as PipeDream explore \textit{asynchronous pipelining}~\cite{harlapPipeline,narayanan2021memoryefficient,yang2020pipemare,gaunt2017ampnet,li2021chimera,shoeybi2019megatron,narayanan2021efficient}, a memory-efficient alternative to standard pipelining. However, these approaches are known to introduce accuracy degradation as they re-order execution stages of the model to minimize memory usage. We do not compare to asynchronous approaches in this paper since accuracy is a critical metric in model selection workloads --- introducing any tradeoff between accuracy and efficiency complicates the objectives of model selection workloads. As such, we only compare to ``exact'' parallel approaches that are mathematically equivalent to standard execution. 

\textbf{Tensor offloading systems} such as SwapAdvisor~\cite{huang2020swapadvisor,meng2017training} enable large models to be trained on a single GPU. Other approaches such as ZeRO-Infinity~\cite{zeroInfinity} and L2L~\cite{l2l} introduce similar designs with further extensions, such as ZeRO-Infinity's CPU-based parameter updates and L2L's Transformer block-swapping design. All of these systems are heavily optimized for single-model execution, where GPU availability for a model is essentially guaranteed across the course of offloaded execution. Hybrid model-task parallelism requires the ability to temporarily ``pause'' model execution partway in favor of a different model. L2L in particular is restrictive in that it only works for Transformers. Both L2L and SwapAdvisor are only capable of using a single GPU and not targeted towards multi-GPU environments. The specialized nature of these designs prevents them from working in the more general context of a multi-model executor, though they can be beneficial for single-model execution. Spilling's flexibility and generality are critical for our hybrid model-task parallelism, and it cannot be replaced by a different offloading design.

\textbf{Parallelization strategy search tools} such as FlexFlow~\cite{flexflow} and Alpa~\cite{alpa} combine a variety of parallel execution strategies using simulators and solvers to identify a near-optimal approach to distributing a model architecture across devices. These approaches do not consider the possibility of task parallelism, instead optimizing each model individually. \system~could potentially be hybridized with these tools in the future to enable more holistic optimization for multi-model workloads, especially in cases where there are more devices than models.

\textbf{Reducing model memory footprints} has received much attention in DL 
systems~\cite{chen2016training, gruslys2016memoryefficient, Kumar:EfficientRematerialization, Jain:Checkmate, TASO}. Model quantization~\cite{jacob2017quantization} in particular has been a popular technique for reducing memory footprints at inference time.
The goal of such systems is \textit{orthogonal} to our own, and memory footprint reduction techniques could be integrated into \system~in the future. One system~\cite{hyptransfer} explores the possibility of transferring hyper-parameter selections from small models to larger models. Our focus is broader, tackling multi-large-model execution in general. Other work on machine teaching~\cite{wang2021gradient} and data distillation~\cite{distillation} aims to minimize the memory footprints of data, but these techniques address a different aspect of memory in DL systems. 

\textbf{Other optimizations for DL systems} that exploit multi-task execution, e.g., systems such as ModelBatch~\cite{modelbatch}, Cerebro~\cite{cerebro:kumar}, SystemML~\cite{systemml}, Krypton~\cite{krypton}, and ASHA~\cite{asha}. ModelBatch raises GPU utilization by altering the DL tool's internal execution kernels. Cerebro proposes a hybrid parallelism scheme named MOP combining task- and data- parallelism, akin to (but different from) SHARP's hybrid model-task parallelism. SystemML also hybridizes task- and data-parallelism, but for classical ML workloads rather than DL. Krypton applies task parallelism to multi-inference workloads with CNNs. ASHA is a new hyperparameter tuning algorithm that accounts for cluster size. None of them tackle larger-than-GPU-memory models, which is our focus. 

\section{Future Work \& Ethical Implications}
\label{sec:improvements}
While \system~already is already the most resource-efficient among the systems we benchmarked on multi-large-model workloads, there are several areas for potential improvement. For example, spilling still has communication latencies. Although this tends to be outweighed by the task parallel speedups that spilling enables, it is can be a bottleneck in some cases. Making use of optimized low-level CUDA data transfer kernels has been shown to improve offloading performance in prior works~\cite{zeroDeep,l2l}. We only use PyTorch-provided communication commands which helps with compatibility as PyTorch develops in the future, but hurts efficiency since they are not optimized for our use-case. 

Another current limitation of \system~is that it inherits one of the restrictions of task parallelism; if there are fewer models than GPUs (e.g. single model training), then the degree of parallelism offered by Shard Alternator Parallelism is bounded by the number of models. Hybridizing with data parallelism, maybe even offloaded data parallelism like ZeRO-3, could enable us to optimize for this setting as well. We leave such complex hybrid-of-hybrid systems to future work.

In the current version of \system~we do not exploit spilling to disk and we do not yet support multi-node environments, which is typically needed for very large datasets. We will explore these optimizations in future work. \system~can also be used for inference with large models (not just training), although we have not explicitly optimized execution for that setting.

By and large, the ethical implications of \system~are mostly positive. We enable scalability for single GPU users, democratize access to large-scale models, and improve efficiency for multi-GPU users. This reduces the energy footprint of model selection, enables faster model development, and encourages replicable model selection practices. That being said, increased accessibility to powerful large-scale DL models must also be paired with increased caution and responsibility.

\section{Conclusion}
Building larger-than-GPU-memory DL models is a pressing need for researchers and practitioners. Such large models are increasingly being deployed in numerous applications outside of technology companies. Unfortunately, the computational demands of such models are impeding rigorous model selection practices such as hyperparameter searches or careful fine-tuning. To tackle this problem, we present \system, a new system for multi-large-model DL training. We present the first-known hybrid of model-parallelism with task-parallelism to enable highly resource-efficient multi-large-model training on a GPU cluster, as well as out-of-the-box model scalability with even just one GPU. Overall, by optimizing multi-large-model workloads more holistically, our work helps make modern DL faster, cheaper, and more accessible to diverse user bases.

\bibliographystyle{plain}
\bibliography{hydra}

\begin{thebibliography}{10}

\bibitem{huggingfaceexample}
Examples.
\newblock {\em Examples - transformers 2.0.0 documentation}.

\bibitem{osbook}
Remzi~H. Arpaci-Dusseau and Andrea~C. Arpaci-Dusseau.
\newblock {\em {Operating Systems: Three Easy Pieces}}.
\newblock Arpaci-Dusseau Books, 1.00 edition, August 2018.

\bibitem{modelparallelism}
Tal Ben{-}Nun and Torsten Hoefler.
\newblock Demystifying parallel and distributed deep learning: An in-depth
  concurrency analysis.
\newblock {\em CoRR}, abs/1802.09941, 2018.

\bibitem{systemml}
Matthias Boehm, Michael~W. Dusenberry, Deron Eriksson, Alexandre~V.
  Evfimievski, Faraz~Makari Manshadi, Niketan Pansare, Berthold Reinwald,
  Frederick~R. Reiss, Prithviraj Sen, Arvind~C. Surve, and Shirish Tatikonda.
\newblock Systemml: Declarative machine learning on spark.
\newblock {\em Proc. VLDB Endow.}, 9(13):1425–1436, sep 2016.

\bibitem{foundationmodels}
Rishi et~al. Bommasani.
\newblock On the opportunities and risks of foundation models.
\newblock 2021.

\bibitem{hellogpt}
Paweł Budzianowski and Ivan Vulić.
\newblock Hello, it's gpt-2 -- how can i help you? towards the use of
  pretrained language models for task-oriented dialogue systems.
\newblock 2019.

\bibitem{chen2016training}
Tianqi Chen, Bing Xu, Chiyuan Zhang, and Carlos Guestrin.
\newblock {T}raining {D}eep {N}ets with {S}ublinear memory cost.
\newblock 2016.

\bibitem{devlinBert}
Jacob Devlin, Ming{-}Wei Chang, Kenton Lee, and Kristina Toutanova.
\newblock {BERT:} pre-training of deep bidirectional transformers for language
  understanding.
\newblock {\em CoRR}, abs/1810.04805, 2018.

\bibitem{vit}
Alexey~Dosovitskiy et~al.
\newblock An image is worth 16x16 words: Transformers for image recognition at
  scale.
\newblock {\em CoRR}, abs/2010.11929, 2020.

\bibitem{huggingface}
Thomas~Wolf et~al.
\newblock Transformers: State-of-the-art natural language processing.
\newblock In {\em Proceedings of the 2020 Conference on Empirical Methods in
  Natural Language Processing: System Demonstrations}, pages 38--45, Online,
  October 2020. Association for Computational Linguistics.

\bibitem{gaunt2017ampnet}
Alexander~L. Gaunt, Matthew~A. Johnson, Maik Riechert, Daniel Tarlow, Ryota
  Tomioka, Dimitrios Vytiniotis, and Sam Webster.
\newblock Ampnet: Asynchronous model-parallel training for dynamic neural
  networks.
\newblock 2017.

\bibitem{rutransformers}
Mikhail Grankin.
\newblock Russian gpt-2.
\newblock {\em GitHub repository}, 2020.

\bibitem{gruslys2016memoryefficient}
Audrūnas Gruslys, Remi Munos, Ivo Danihelka, Marc Lanctot, and Alex Graves.
\newblock {M}emory-{E}fficient {B}ackpropagation {T}hrough {T}ime.
\newblock 2016.

\bibitem{gurobi}
LLC Gurobi~Optimization.
\newblock {G}urobi {O}ptimizer {R}eference {M}anual.
\newblock 2021.

\bibitem{harlapPipeline}
Aaron Harlap, Deepak Narayanan, Amar Phanishayee, Vivek Seshadri, Nikhil~R.
  Devanur, Gregory~R. Ganger, and Phillip~B. Gibbons.
\newblock Pipedream: Fast and efficient pipeline parallel {DNN} training.
\newblock {\em CoRR}, abs/1806.03377, 2018.

\bibitem{deepspeedissue1}
Daniel Hesslow.
\newblock Module names gpt2modelpipe \& paralleltransformerlayerpipe is
  hardcoded in deepspeed.
\newblock {\em GitHub Issue}, 2021.

\bibitem{huang2020swapadvisor}
Chien-Chin Huang, Gu~Jin, and Jinyang Li.
\newblock Swapadvisor: Pushing deep learning beyond the gpu memory limit via
  smart swapping.
\newblock In {\em Proceedings of the Twenty-Fifth International Conference on
  Architectural Support for Programming Languages and Operating Systems}, pages
  1341--1355, 2020.

\bibitem{gpipe}
Yanping Huang, Yonglong Cheng, Dehao Chen, HyoukJoong Lee, Jiquan Ngiam,
  Quoc~V. Le, and Zhifeng Chen.
\newblock {GP}ipe: {E}fficient {T}raining of {G}iant {N}eural {N}etworks using
  {P}ipeline parallelism.
\newblock {\em CoRR}, abs/1811.06965, 2018.

\bibitem{jacob2017quantization}
Benoit Jacob, Skirmantas Kligys, Bo~Chen, Menglong Zhu, Matthew Tang, Andrew
  Howard, Hartwig Adam, and Dmitry Kalenichenko.
\newblock Quantization and training of neural networks for efficient
  integer-arithmetic-only inference.
\newblock 2017.

\bibitem{Jain:Checkmate}
Paras Jain, Ajay Jain, Aniruddha Nrusimha, Amir Gholami, Pieter Abbeel, Kurt
  Keutzer, Ion Stoica, and Joseph~E. Gonzalez.
\newblock Checkmate: Breaking the memory wall with optimal tensor
  rematerialization.
\newblock {\em CoRR}, abs/1910.02653, 2019.

\bibitem{TASO}
Zhihao Jia, Oded Padon, James Thomas, Todd Warszawski, Matei Zaharia, and Alex
  Aiken.
\newblock {TASO}: {O}ptimizing {D}eep {L}earning {C}omputation with {A}utomatic
  {G}eneration of {G}raph {S}ubstitutions.
\newblock {\em SOSP '19}, 2019.

\bibitem{flexflow}
Zhihao Jia, Matei Zaharia, and Alex Aiken.
\newblock {B}eyond {D}ata and {M}odel {P}arallelism for {D}eep {N}eural
  {N}etworks.
\newblock {\em CoRR}, 2018.

\bibitem{torchgpipe}
Chiheon Kim, Heungsub Lee, Myungryong Jeong, Woonhyuk Baek, Boogeon Yoon, Ildoo
  Kim, Sungbin Lim, and Sungwoong Kim.
\newblock torchgpipe: On-the-fly pipeline parallelism for training giant
  models.
\newblock 2020.

\bibitem{cerebro:kumar}
Arun Kumar, Supun Nakandala, Yuhao Zhang, Side Li, Advitya Gemawat, and Kabir
  Nagrecha.
\newblock Cerebro: {A} layered data platform for scalable deep learning.
\newblock In {\em 11th Conference on Innovative Data Systems Research, {CIDR}
  2021, Virtual Event, January 11-15, 2021, Online Proceedings}.
  www.cidrdb.org, 2021.

\bibitem{Kumar:EfficientRematerialization}
Ravi Kumar, Manish Purohit, Zoya Svitkina, Erik Vee, and Joshua Wang.
\newblock {E}fficient {R}ematerialization for {D}eep {N}etworks.
\newblock 32:15172--15181, 2019.

\bibitem{asha}
Liam Li, Kevin Jamieson, Afshin Rostamizadeh, Ekaterina Gonina, Moritz Hardt,
  Benjamin Recht, and Ameet Talwalkar.
\newblock A system for massively parallel hyperparameter tuning.
\newblock 2020.

\bibitem{li2021chimera}
Shigang Li and Torsten Hoefler.
\newblock Chimera: efficiently training large-scale neural networks with
  bidirectional pipelines.
\newblock In {\em Proceedings of the International Conference for High
  Performance Computing, Networking, Storage and Analysis}, pages 1--14, 2021.

\bibitem{meng2017training}
Chen Meng, Minmin Sun, Jun Yang, Minghui Qiu, and Yang Gu.
\newblock Training deeper models by gpu memory optimization on tensorflow.
\newblock In {\em Proc. of ML Systems Workshop in NIPS}, 2017.

\bibitem{wikitext2}
Stephen Merity, Caiming Xiong, James Bradbury, and Richard Socher.
\newblock Pointer sentinel mixture models.
\newblock {\em CoRR}, abs/1609.07843, 2016.

\bibitem{hydra_cerebro}
Kabir Nagrecha.
\newblock Model-parallel model selection for deep learning systems.
\newblock page 2929–2931, 2021.

\bibitem{krypton}
Supun Nakandala, Kabir Nagrecha, Arun Kumar, and Yannis Papakonstantinou.
\newblock Incremental and approximate computations for accelerating deep cnn
  inference.
\newblock {\em ACM Trans. Database Syst.}, 45(4), December 2020.

\bibitem{narayanan2021memoryefficient}
Deepak Narayanan, Amar Phanishayee, Kaiyu Shi, Xie Chen, and Matei Zaharia.
\newblock Memory-efficient pipeline-parallel dnn training.
\newblock 2021.

\bibitem{modelbatch}
Deepak Narayanan, Keshav Santhanam, and Matei Zaharia.
\newblock Accelerating model search with model batching.
\newblock {\em Proceedings of Fourth Conference on Machine Learning and Systems
  (MLSys'18)}, 2018.

\bibitem{narayanan2021efficient}
Deepak Narayanan, Mohammad Shoeybi, Jared Casper, Patrick LeGresley, Mostofa
  Patwary, Vijay Korthikanti, Dmitri Vainbrand, Prethvi Kashinkunti, Julie
  Bernauer, Bryan Catanzaro, Amar Phanishayee, and Matei Zaharia.
\newblock {E}fficient {L}arge-{S}cale {L}anguage {M}odel {T}raining on {GPU}
  {C}lusters.
\newblock {\em CoRR}, abs/2104.04473, 2021.

\bibitem{l2l}
Bharadwaj Pudipeddi, Maral Mesmakhosroshahi, Jinwen Xi, and Sujeeth Bharadwaj.
\newblock Training large neural networks with constant memory using a new
  execution algorithm.
\newblock 2020.

\bibitem{radford2019language}
Alec Radford et~al.
\newblock Language models are unsupervised multitask learners.
\newblock {\em OpenAI blog}, 1(8):9, 2019.

\bibitem{zeroInfinity}
Samyam Rajbhandari, Olatunji Ruwase, Jeff Rasley, Shaden Smith, and Yuxiong He.
\newblock Zero-infinity: Breaking the gpu memory wall for extreme scale deep
  learning.
\newblock 2021.

\bibitem{zeroDeep}
Samyam Rajbhari, Jeff Rasley, Olatunji Ruwase, and Yuxiong He.
\newblock {Z}e{RO}: {M}emory {O}ptimizations {T}oward {T}raining {T}rillion
  {P}arameter {M}odels.
\newblock 2020.

\bibitem{zeroOpt}
Jie Ren, Samyam Rajbhandari, Reza~Yazdani Aminabadi, Olatunji Ruwase, Shuangyan
  Yang, Minjia Zhang, Dong Li, and Yuxiong He.
\newblock Zero-offload: Democratizing billion-scale model training.
\newblock 2021.

\bibitem{gpt2portugal}
Elisa Terumi~Rubel Schneider, João Vitor~Andrioli de~Souza, Yohan~Bonescki
  Gumiel, Claudia Moro, and Emerson~Cabrera Paraiso.
\newblock A gpt-2 language model for biomedical texts in portuguese.
\newblock In {\em 2021 IEEE 34th International Symposium on Computer-Based
  Medical Systems (CBMS)}, pages 474--479, 2021.

\bibitem{deepspeedissue2}
sdtblck.
\newblock Activation checkpointing breaks for some layers in pipelinemodule.
\newblock {\em GitHub repository}, 2021.

\bibitem{understandingml}
S.~Shaleve-Shwartz and S.~Ben-David.
\newblock {\em {Understanding Machine Learning: from Theory to Algorithms}}.
\newblock {Cambridge University Press}, 2014.

\bibitem{shoeybi2019megatron}
Mohammad Shoeybi, Mostofa Patwary, Raul Puri, Patrick LeGresley, Jared Casper,
  and Bryan Catanzaro.
\newblock Megatron-lm: Training multi-billion parameter language models using
  model parallelism.
\newblock {\em CoRR}, abs/1909.08053, 2019.

\bibitem{sohonilowmem}
Nimit~Sharad Sohoni, Christopher~Richard Aberger, Megan Leszczynski, Jian
  Zhang, and Christopher R{\'{e}}.
\newblock Low-memory neural network training: {A} technical report.
\newblock {\em CoRR}, abs/1904.10631, 2019.

\bibitem{brown2020language}
et~al. Tom B.~Brown.
\newblock {L}anguage {M}odels are {F}ew-{S}hot {L}earners.
\newblock {\em CoRR}, abs/2005.14165, 2020.

\bibitem{wang2019cas}
Chenguang Wang, Mu~Li, and Alexander~J. Smola.
\newblock Language models with transformers.
\newblock 2019.

\bibitem{wang2021gradient}
Pei Wang, Kabir Nagrecha, and Nuno Vasconcelos.
\newblock Gradient-based algorithms for machine teaching.
\newblock In {\em Proceedings of the IEEE/CVF Conference on Computer Vision and
  Pattern Recognition}, pages 1387--1396, 2021.

\bibitem{distillation}
Tongzhou Wang, Jun{-}Yan Zhu, Antonio Torralba, and Alexei~A. Efros.
\newblock {Dataset Distillation}.
\newblock {\em CoRR}, abs/1811.10959, 2018.

\bibitem{yang2020pipemare}
Bowen Yang, Jian Zhang, Jonathan Li, Christopher Ré, Christopher~R. Aberger,
  and Christopher~De Sa.
\newblock Pipemare: Asynchronous pipeline parallel dnn training.
\newblock 2020.

\bibitem{hyptransfer}
Greg Yang, Edward~J. Hu, Igor Babuschkin, Szymon Sidor, Xiaodong Liu, David
  Farhi, Nick Ryder, Jakub Pachocki, Weizhu Chen, and Jianfeng Gao.
\newblock Tensor programs v: Tuning large neural networks via zero-shot
  hyperparameter transfer.
\newblock 2022.

\bibitem{tabert}
Pengcheng Yin, Graham Neubig, Wen{-}tau Yih, and Sebastian Riedel.
\newblock Tabert: Pretraining for joint understanding of textual and tabular
  data.
\newblock {\em CoRR}, abs/2005.08314, 2020.

\bibitem{alpa}
Lianmin Zheng, Zhuohan Li, Hao Zhang, Yonghao Zhuang, Zhifeng Chen, Yanping
  Huang, Yida Wang, Yuanzhong Xu, Danyang Zhuo, Joseph~E Gonzalez, et~al.
\newblock Alpa: Automating inter-and intra-operator parallelism for distributed
  deep learning.
\newblock {\em arXiv preprint arXiv:2201.12023}, 2022.

\end{thebibliography}

\end{document}